\documentclass[pra, reprint,superscriptaddress]{revtex4-1}
\usepackage{amsmath}
\usepackage{amsfonts}
\usepackage{graphicx}
 \usepackage{physics}
\usepackage[usenames]{color}

\usepackage{braket}
\usepackage{comment}

\begin{document}

\author{Sayan Choudhury}
\email{sc2385@cornell.edu}
\affiliation{Department of Physics and Astronomy, University of Pittsburgh, PA 15260, USA}

\author{Kazi R. Islam}
\thanks{Current Affiliation: Department of Physics, University of Minnesota Twin Cities, MN 55455, USA}
\affiliation{Department of Physics and Astronomy, Rice University, Houston, Texas 77251, USA}

\author{Yanhua Hou} 
\affiliation{Department of Physics and Astronomy, Rice University, Houston, Texas 77251, USA}

\author{Jim A. Aman}
\affiliation{Department of Physics and Astronomy, Rice University, Houston, Texas 77251, USA}

\author{Thomas C. Killian}
\affiliation{Department of Physics and Astronomy, Rice University, Houston, Texas 77251, USA}

\author{Kaden R. A. Hazzard}
\email{kaden@rice.edu}
\affiliation{Department of Physics and Astronomy, Rice University, Houston, Texas 77251, USA}

\title{Collective modes of ultracold fermionic alkaline-earth gases with SU($N$) symmetry}
\date{\today}

\begin{abstract}
We calculate the collective modes  of ultracold trapped alkaline-earth fermionic atoms, which possess an SU($N$) symmetry of the nuclear spin degree of freedom, and a controllable $N$, with $N$ as large  as $10$.    We  calculate the breathing and quadrupole modes of two-dimensional and three-dimensional harmonically trapped gases in the normal phase. We particularly concentrate on two-dimensional gases, where the shift is more  accessible experimentally, and the physics has special features. We present results as a function of temperature, interaction strength, density, and $N$. We include calculations across the collisionless to hydrodynamic crossover.  We assume the gas is interacting weakly, such that it can be described by a  Boltzmann-Vlasov equation that includes both mean-field terms and the collision integral. We solve this with an approximate scaling ansatz, taking care in two-dimensions to preserve the scaling symmetry of the system. We predict the collective mode frequency shifts and damping, showing that these are measurable in experimentally relevant regimes. We expect these results  to furnish powerful tools to characterize interactions and the state of alkaline-earth gases,  as well as to lay the foundation for future work, for example on strongly interacting gases and SU($N$) spin modes.  
\end{abstract}

\maketitle

\section{Introduction}  
Ultracold alkaline-earth-like atoms such as Yb and Sr have unique  properties that open  new regimes of many-body physics~\cite{daley:quantum_2011,stellmer:degenerate_2014,cazalilla:ultracold_2014,he:recent_2019}. One example is that their closed-shell electronic structure provides  a long-lived clock state that has enabled  optical clocks with a precision approaching $10^{-19}$. Another example is the fermionic isotopes' large nuclear spin $I$, leading to a large number  $N=2I+1$ of degenerate internal states on each atom, where any $N$ can be produced up to $N=6$ (in Yb) and  $N=10$ (in Sr). Equally important to the large degeneracy is the  SU($N$) symmetry that interactions between the atoms enjoy~\cite{wu:exact_2003,cazalilla:utracold_2009,gorshkov:two-orbital_2010,stellmer:detection_2011,zhang:spectroscopic_2014}.  
Although one might naively expect that such large  spins become classical, it is known that in some circumstances the large symmetry group can enhance quantum fluctuations such that they remain relevant even as $N\rightarrow\infty$, and that such fluctuations give rise to exotic phenomena such as chiral spin liquids~\cite{hermele:mott_2009,hermele:topological_2011}, molecular Luttinger liquids, symmetry protected topological phases, quantum liquids,  valence bond solid states, and magnetically ordered states~\cite{honerkamp:ultracold_2004,gorelik:mott_2009,toth:three-sublattice_2010,rapp:ground-state_2011,manmana:sun_2011,corboz:simultaneous_2011,hazzard:high-temperature_2012,bonnes:adiabatic_2012,messio:entropy_2012,bauer:three-sublattice_2012,inaba:superfluid_2013,sotnikov:magnetic_2014,nataf:exact_2014,sotnikov:critical_2015,nataf:plaquette_2016,chen:synthetic_2016,capponi:phases_2016,nataf:chiral_2016}, which are beginning to be explored experimentally~\cite{hofrichter:direct_2016,taie:su6_2012,ozawa:antiferromagnetic_2018}. In light of this, it is especially interesting to explore how the physics depends on $N$. 

The properties of  interacting Fermi gases are broadly studied, and 
two-dimensional (2D) gases with short-ranged interactions are particularly interesting for two reasons~\cite{levinsen:strongly_2015}. The first is that their reduced dimensionality enhances  quantum and thermal fluctuations, limiting the applicability of mean-field theory. The second is that they possess intriguing special features: an SO(2,1) scaling symmetry at the classical level that is broken by a quantum anomaly for a variety of bosonic and fermionic systems~\cite{pitaevskii:breathing_1997,werner:unitary_2006,olshanii:example_2010,hofmann:quantum_2012,gao:breathing_2012,moroz:scale_2012,taylor:apparent_2012,chafin:scale_2013,ordonez:path-integral_2016,daza:virial_2018},
and recently-predicted long-lived memory effects in homogeneous systems~\cite{ledwith:hierarchy_2019}.

Collective modes -- macroscopic oscillations (possibly damped) of a trapped system in response to an external perturbation -- are a powerful probe of matter. They reveal information about the equation of state and quasiparticle properties, especially the quasiparticle collisions.  They have therefore been central to   experiments studying  ultracold matter. 
The collective breathing (i.e., monopole) and quadrupole density modes   have been measured in 2D spin-1/2 SU(2) Fermi gases~\cite{vogt:scale_2012,schaefer:shear_2012,enss:shear_2012,holten:anomalous_2018,peppler:quantum_2018}, and have spurred a variety of theoretical explorations~\cite{bruun:shear_2012,baur:collective_2013,Chiacchiera:damping_2013,mulkerin:collective_2018,hu:reduced_2019}.
Working in 2D is  also  beneficial for measuring collective modes of alkaline-earth-like gases in experiment. The reason is that  confining the system in the third dimension increases the effective interactions strength, and thus increases the collective mode frequency shifts and damping rates. This is especially important since alkaline earth atoms have no ground state magnetic Feshbach resonances. 
 
Given their fundamental interest and accessibility, it is   interesting to study the collective modes of 2D SU($N$) Fermi gases.  
Their behavior includes the interesting behavior of spin-1/2 SU(2) Fermi gases as a limiting case, but goes beyond this  with   an additional control parameter $N$. Changing $N$ may, for example,  tune  the strength of quantum fluctuations.  
Moreover, SU($N$) gases will also display collective oscillations of the spin degrees of freedom. These are a richer analog of the spin modes measured for $N=2$ Fermi gases in Ref.~\cite{koschorreck:universal_2013,bardon:transverse_2014,trotzky:observation_2015,valtolina:exploring_2017,enss:universal_2019},  which have shed light on correlated quantum transport, for example suggesting fundamental quantum bounds on hydrodynamic transport coefficients. Although we focus in this paper  on density rather than spin modes, the paper also sets up a theoretical framework for treating the latter.
Initial measurements of density collective modes have been performed for  SU($N$) Fermi gases in 1D~\cite{pagano:one-dimensional_2014}, finding strongly correlated states through a crossover from non-interacting fermions at $N=1$ to nearly bosonic behavior at $N=10$.

 In this paper, we calculate the collective mode frequencies and damping rates in a weakly interacting two-dimensional (2D) SU($N$) Fermi gas as a function of interaction strength, temperature, and $N$.  We  focus  on the breathing and quadrupole density modes.
 In addition to treating the weakly interacting situation, we expect the theory developed here to lay the groundwork  to explore strongly-interacting 2D alkaline-earth-like atom gases, and associated questions of spin modes and spin transport. We also briefly consider the three-dimensional (3D) gas in Sec.~\ref{sec:3D}. Excitingly, as the present manuscript was being finalized, Ref.~\cite{he:collective_2019} has measured the breathing mode and quadrupole mode frequencies and damping in a 2D SU($N$) Fermi gas for $N=1,2,\ldots,6$ in the collisionless limit. We will discuss these experiments in comparison with our calculations (along with other experimental predictions) in  Sec.~\ref{sec:experiment}.
  
While our focus is on
SU($N$) 
Fermi gases, several of our results are also useful for spin-1/2 SU($2$) gases, as occur in experiments with ultracold alkali atoms. 
It is worth emphasizing two results in this regard. First, our approximations are  carefully designed to ensure consistency with the subtle  SO(2,1)  scaling symmetry enjoyed by the system at the classical level (i.e. in the absence of the quantum anomaly). This symmetry is implemented consistently even in the presence of important mean-field shifts, where common alternative techniques break the symmetry and would give physically incorrect results. Second, our approximations are flexible enough to capture the the quadrupole collective mode frequency  in the collisionless limit arising from  mean-field interaction effects 
We will discuss how our results compare  to those obtained from alternative popular approximations that fail to have these desirable properties in Sec.~\ref{sec:coll-freq}.

 Section~\ref{sec:methods} describes the experimental system we consider, the  collective modes we focus on, and  the theoretical framework and approximations we use to describe the nonequilibrium dynamics (Boltzmann equation with a scaling ansatz solution). Section~\ref{sec:results} presents our results for the collective modes. 
 It describes the dependence of collective mode frequencies and damping on system parameters: interaction strength $g$, $N$, temperature $T$, number of particles $N_p$, and radial  and transverse trap frequencies, $\omega_{\text{tr}}$ and $\omega_z$, respectively. In particular, Sec.~\ref{sec:experiment} evaluates these shifts and dampings for typical experimental parameters, and compares to very recently obtained measured SU($N$) collective mode properties in the collisionless limit~\cite{he:collective_2019}.  Sec.~\ref{sec:3D} briefly presents results for 3D gases. Section~\ref{sec:conclusions} concludes and provides an outlook.

\begin{figure}
\includegraphics[scale=.45]{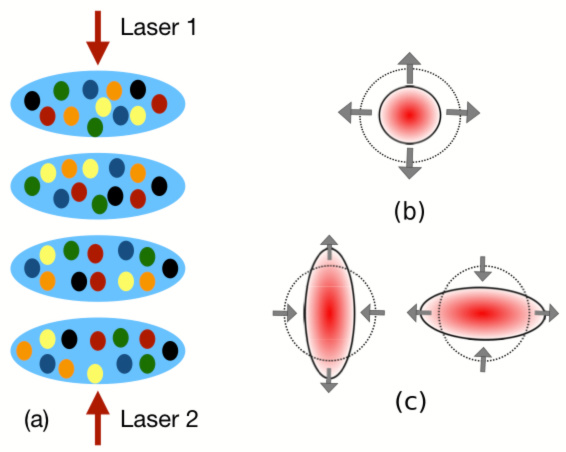}
\caption{Collective modes of a 2D SU($N$) Fermi gases. (a) Fermionic alkaline-earth atoms can be confined to a single 2D layer or to an array of 2D layers formed by an optical lattice (depicted). Collective modes may be excited, for example by suddenly changing a trap frequency. (b-c) The real space deformations corresponding to the three lowest angular momentum modes, the breathing mode (b), dipole mode (not depicted), and quadrupole mode (c).  
\label{fig:cartoon}}
\end{figure}

 \section{Collective modes and theoretical methods \label{sec:methods}}

 2D gases  can be experimentally realized in ultracold alkaline-earth-like atoms  
 by directly confining them to  a single layer (for example with evanescent fields), or by creating an array of 2D systems via  a one dimensional optical lattice, as illustrated in Fig.~\ref{fig:cartoon}(a). We will assume that the lattice is deep enough that the 2D layers are uncoupled. Also, we will assume an isotropic harmonic trap potential with trap frequency $\omega_{\text{tr}}$. 
 
 Collective modes may be excited by suddenly changing system parameters. For  collective modes of the density, as we consider in this paper, this is often done by suddenly changing a trap frequency by a small amount ($\sim~\!10\%)$. Which modes are excited will depend on the symmetry of this perturbation (and of the original trap). A generic perturbation excites a superposition of modes of different symmetry, but often experiments choose perturbations  to couple to modes with a single symmetry. For example,  an isotropic change of trap frequencies in an isotropic trap excites only  breathing modes,  illustrated in Fig.~\ref{fig:cartoon}(b). The other  mode we consider in this paper is the quadrupole mode is illustrated in Fig.~\ref{fig:cartoon}(c).  To measure these modes, experiments can track \textit{in situ} oscillations of the density profile, or oscillations in time-of-flight, which measure the momentum distribution. Sometimes, easier-to-access observables are measured as proxies, for example loss as a function of time, which allows one  to measure the frequencies and damping rates, although it provides less details about the spatial and momentum-space mode structure.  
Recently, He~\textit{et al.} have measured the breathing and quadrupole collective mode frequencies and damping times in SU($N$) Yb gases confined in a one-dimensional lattice, for various $N$ up to $N=6$~\cite{he:collective_2019}.\\

  A 2D alkaline-earth fermionic gas with $N_\text{p}$ particles can be described by the grand canonical Hamiltonian 
\begin{align}
 \label{Hamiltonian}
 H&=\sum_{\alpha}\int \! d^2 \mathbf{r}\, \psi^{\dagger}_\alpha(\mathbf{r})\left(-\frac{\hbar^2}{2m}\nabla^2-\mu+V(r)\right)\psi_\alpha(\mathbf{r}) \nonumber \\
&{} +\frac{g_{2D}}{2} \!\!\sum_{\alpha \neq \beta}\int\! d^2 \mathbf{r}\, \psi^{\dagger}_\alpha(\mathbf{r})\psi^{\dagger}_\beta(\mathbf{r})\psi^{\phantom{\dagger}}_\beta(\mathbf{r})\psi_\alpha^{\phantom{\dagger}}(\mathbf{r}),
  \end{align}
   where $\psi^{\dagger}_\alpha(\mathbf{r})$ is the fermionic creation operator creating an atom at  position $\mathbf{r}$ with spin index $\alpha=1,\ldots,N$, $\mu$ is the chemical potential, $m$ is the mass of the atom, $V(r)=m\omega_{\text{tr}}^2 (x^2+y^2)/2$ is the harmonic trap potential with frequency $\omega_{\text{tr}}$, and $g_{2D}$ is the interaction strength. 
In principle, this contact interaction must be regularized, but at the level of approximations we use throughout, this will be unnecessary. For a sufficiently deep lattice, so that the potential confining the atoms to the 2D plane can be described by an additive potential $m \omega_z^2 z^2/2$ (with $z$ the displacement perpendicular to the plane),  the coupling constant is~\cite{petrov:interatomic_2001}  
\begin{equation}
   g_{2D}= \frac{ 2\pi \hbar^2}{m \ln(q a_{\text{2D}})},
\end{equation}
   where $a_{\text{2D}} = l_z\sqrt{\pi/B} \exp(-\sqrt{\pi/2} l_z/a_{\text{3D}})$, $B=0.915$ $a_{\text{3D}}$ is the three-dimensional $s$-wave scattering length, $l_z=\sqrt{\hbar/(m\omega_z)}$, and $q \sim \sqrt{n}$ is a characteristic momentum that determines the density dependent coupling ($n$ being the total density). The momentum factor, $q$ is the Fermi momentum, $k_F$ \cite{ghosh2002splitting}, and the de Broglie wavelength, $\sqrt{mT}/\hbar$ in the low and high temperature limit respectively \cite{petrov:interatomic_2001}.  \\

   We calculate the collective modes by employing two approximations, which are reasonable in the limits considered in this paper. First, we assume that the system is weakly interacting,  $1/{\ln{(1/n a_{\text{2D}}^2)} \ll 1}$ (where $n$ is the total density), its temperature is sufficiently high such there is no pairing, and that the length scales over which there is any spatial coherence is small compared to the trap size (though the gas may still be deeply degenerate). In practice this means that the system must be well above the superconducting transition temperature and the length scale on which the collective modes vary must be long compared to the thermal de Broglie wavelength.  This allows us to describe the system's dynamics by a Boltzmann equation, including mean-field interactions in addition to the collision integral, which governs the phase space distribution function (defined later). The resulting Boltzmann-Vlasov (BV) equation ~\footnote{We use this to refer to the Boltzmann equation with both mean field interactions and the collision integral, but we note that some authors use this terminology to refer strictly to the collisionless equation.} is a 4+1 dimensional partial differential-integral equation, and as such would be extremely  demanding to solve numerically. 
   
   Second, we approximate the collision integral with a  relaxation time approximation. This is an uncontrolled, but standard, approach to calculating transport and collective modes within a Boltzmann equation framework.  This approximation will be explained in detail in Sec.~\ref{sec:results}.

   To solve the BV equation in the relaxation time approximation, we assume an ansatz for the phase space distribution function. The ansatz is carefully constructed to respect the SO(2,1) symmetry of the system, while simultaneously being flexible enough to capture the collective modes' shifts and damping. As shown in Ref.~\cite{pitaevskii:breathing_1997}, this ansatz provides an exact solution to the collective modes of the BV equation in 2D, when the confinement is isotropic.

   Under these assumptions, it is valid to apply mean-field theory to the Hamiltonian in Eq.~\eqref{Hamiltonian} in the density channel. We assume that the density of each species is the same ($n_{\alpha}=n_{\beta}=n_0$, for all $\{\alpha,\beta \} \in \{1,\ldots N\}$), where $n_0$ is the density of each species. [This holds  in any state that  preserves the Hamiltonian's SU($N$) symmetry.] The mean-field Hamiltonian is then
   \begin{equation}
   \label{Mean field}
   H_{\text{MF}}=\sum_{\alpha}\!\int\! d^2 \mathbf{r}\, \psi^{\dagger}_\alpha(\mathbf{r})\bigg(\!-\frac{\hbar^2}{2m}\nabla^2-\mu  
   +V(r)+U_{\text{MF}} \!\bigg)\psi_\alpha^{\phantom \dagger}(\mathbf{r}),
   \end{equation}
   up to irrelevant constants, and 
   \begin{equation}
\label{MF potential}
U_{\text{MF}}= g_{\text{2D}}\frac{(N-1)}{N}n^{\text{tot}}(\mathbf{r}) 
\end{equation} 
where 
$n^{\text{tot}}(\mathbf{r})=\sum_{\alpha=1}^N n_{\alpha}(\mathbf{r})=N n_0(\mathbf{r})$ is the density of the gas at position $\mathbf{r}$. The chemical potential $\mu$ is chosen to give the total number of particles $N_{p}$ by
\begin{equation}
        N_p=\int\! d^2\mathbf{r} \,n^{\text{tot}}(\mathbf{r}) =N \int \! d^2\mathbf{r}\, n_0(\mathbf{r})
    \label{total particle}
\end{equation}     

To calculate the collective mode dynamics, we will use the BV kinetic equation, which is a semi-classical method, solved by a scaling ansatz, and linearize for small displacements from equilibrium. The  BV kinetic equation is accurate when the conditions outlined above of weak interactions and  $k_\text{B}T \gg \hbar \omega_{\text{tr}}$ are satisfied. In this limit, we assume that the effects of quantum interference can be neglected, and quasi-particles are well-defined. Furthermore, since our initial state and Hamiltonian are invariant under SU($N$) symmetry (i.e. we are studying the density modes), we have assumed that each spin component is described by the same  semi-classical distribution function $f(\mathbf{r},\mathbf{p},t)$
which satisfies BV kinetic equation, 
\begin{equation}
\frac{\partial f}{\partial t}+ 
\frac{\textbf{p}}{m}\cdot\frac{\partial f}{\partial \textbf{r}}-\frac{\partial}{\partial\mathbf{r}}\left(V(\mathbf{r})+U_{\text{MF}}(\mathbf{r})\right)\cdot\frac{\partial f}{\partial \textbf{p}}=-I_{\text{coll}}[f],
\label{BV}
\end{equation}
where $I_{\text{coll}}[f]$ is the collision integral, and $U_{\text{MF}}$ is the mean-field interaction energy. The mean field interaction energy is given by
Eq.~\eqref{MF potential}
where the density $n^{\text{tot}}(\mathbf{r})$ is determined from the distribution function by
\begin{equation}
\label{total denisty}
n^{\text{tot}}(\mathbf{r})=N \int\! \frac{d^2\mathbf{k}}{(2 \pi)^2} f(\mathbf{r},\mathbf{k}),
\end{equation}
 and the collision integral, $I_{\text{coll}}[f]$ is 
\begin{align}
\label{colleq}
    I_{\text{coll}} [f_1]&=\int\!\frac{d^2 p_2}{(2 \pi \hbar)^2}\frac{m \hbar}{4 \pi} \int_0^{2 \pi}  \! d\theta |\mathcal{T} (|{\mathbf{p}_r}|)|^2 (N-1)  \nonumber \\
    &{}\times\left[f_1 f_2 (1-f_3)(1-f_4) - f_3 f_4 (1-f_1)(1-f_2)\right], \nonumber\\
\end{align}
where we define $f_j=f(\mathbf{r},\mathbf{p}_j)$, and $\mathbf{p}_3$ and $\mathbf{p}_4$ are given in terms of $\mathbf{p}_1$, $\mathbf{p}_2$, and $\theta$ as follows: $\theta$ is the angle between the outgoing relative angular momentum $\mathbf{p}_r = (\mathbf{p}_3 - \mathbf{p}_4)/2$ and the center of mass momentum, ($\mathbf{p}_1+\mathbf{p}_2$), $\mathbf{p}_4$ is given by conservation of the center-of-mass momentum $\mathbf{p}_3+\mathbf{p}_4=\mathbf{p}_1+\mathbf{p}_2$, and by the conservation of energy we obtain $|\mathbf{p}_r| = |\mathbf{p}_r^{\prime}|$, where $\mathbf{p}_r^{\prime}=(\mathbf{p}_1 - \mathbf{p}_2)/2$. The low energy $T$-matrix describing the collision between two atoms with different spins in 2D (in the vacuum) is given by \cite{landau2013quantum}:
\begin{equation}
    \mathcal{T} (q) = \frac{4 \pi}{m} \frac{1}{\ln{(1/q^2 a_{\text{2D}}^2)}+i \pi}.
\end{equation}
 
It is hard to solve the BV equation exactly, even numerically, since this is 5 dimensional partial integro-differential equation. In this paper, we employ a scaling ansatz for $f(\mathbf{r},\mathbf{p},t)$ \cite{guery-odelin:mean-field_2002,pedri2003dynamics,dong2015transition,wachtler2017low}. Our ansatz $f_{sc}$ is defined as
\begin{equation}
\label{scaling}
f_{sc}(\mathbf{r},\mathbf{v},t) =\Gamma(t) f^0(\mathbf{R}(t),\mathbf{V}(t)) 
\end{equation}
where $R_i(t) =\frac{r_i}{b_i(t)}, V_i  =\frac{1}{\sqrt{\theta_i(t)}} \left(v_i-\frac{\dot{b}_i(t)}{b_i(t)} r_i \right)$, and 
$\Gamma(t) =\frac{1}{\prod_{i=1}^2 b_i(t)\sqrt{\theta_i(t)}}$, where $b_i$ and  $\theta_i$ are functions of time that will be determined to give the best solution to the BV equation.
The equilibrium distribution function,  $f^0$, is defined by $I_{\text{coll}}[f^0]=0$, which gives 
\begin{equation}
\label{equilibrium}
m\textbf{v}\cdot\frac{\partial f^0}{\partial \textbf{r}}=\frac{\partial}{\partial \mathbf{r}}(V(r)+U_{\text{MF}})\cdot\frac{\partial f^0}{\partial \textbf{v}}.
\end{equation} As  mentioned, the scaling ansatz respects the classical SO$(2,1)$ scaling symmetry. Quantum effects can lead to a breaking of this symmetry and an anomalous correction to the breathing mode frequency. A calculation of this quantum anomaly is beyond the scope of the BV equation, and we restrict our calculations to the regime where the BV equation is valid.

Additionally, we treat the collision integral in the relaxation time approximation,
\begin{equation}
\label{relaxation time apprx}
I_{\text{coll}}[f]=\frac{f-f^0}{\tau}, 
\end{equation}
where $\tau$ is the relaxation time of the collision, which is calculated in Sec.~\ref{sec:-compute-xi-tau}.  

 \section{Results \label{sec:results}} 

 In this section, we will compute the collective mode frequencies (Sec.~\ref{sec:coll-freq}) and damping rates (Sec.~\ref{sec:-compute-xi-tau}) of the SU($N$) Fermi gas. We find that the breathing mode frequency has no dependence on the interaction strength, reflecting the SO(2,1) symmetry of the system at the classical level. However, the quadrupole mode frequencies exhibit an interaction dependent shift, and damping. We discuss how this shift and damping rates depend on $N$, and estimate the values of these quantities for reasonable experimental parameters.

 \subsection{Collective mode frequencies \label{sec:coll-freq}}
 
 Following Ref.~\cite{pedri2003dynamics}, we compute the average moments of $R_iV_i$ and $V_i^2$ and obtain the following equations for $b_i$ and $\theta_i$:
\begin{align}
\label{bequation}
\ddot{b_i}+\omega_{\text{tr}}^2 b_i-\omega_{\text{tr}}^2 \frac{\theta_i}{b_i}+\omega_{\text{tr}}^2 \xi \left(\frac{\theta_i}{b_i}-\frac{1}{b_i \prod_j b_j}\right)&=0 \\
\label{thetaequ}
\dot{\theta_i}+2\frac{\dot{b}_i}{b_i}\theta_i=-\frac{\theta_i-\bar{\theta}}{\tau}&,
\end{align}
where $i \in \{x,y\}$, $\bar{\theta}=\frac{1}{2}\sum_{i}\theta_i$, and
\begin{equation}
\label{shifteqn}
\xi = \frac{\langle U_{\text{MF}} \rangle}{\langle  m \omega_{\text{tr}}^2 (x^2 + y^2) \rangle}
\end{equation} where $\langle \cdots \rangle=\int\! d^2\mathbf{r} \!\int \!\frac{d^2 \mathbf{k}}{\left( 2\pi\right)^2} f^0\left( \cdots \right)$. For the remainder of this section, we give our results in terms of $\xi$ and $\tau$. We will then calculate $\xi$ and $\tau$ in terms of system parameters in Sec.~\ref{sec:-compute-xi-tau}.

 We linearize Eqs.~\eqref{bequation} and~\eqref{thetaequ} around the equilibrium values $\left(b_i=1+\delta b_i, \theta_i=1+\delta \theta_i \right)$ to get the collective mode frequencies of the density oscillations. We obtain
\begin{align}
&\ddot{\delta b}_j+\omega_{\text{tr}}^2(2+\xi)\delta b_j+ \xi \omega_{\text{tr}}^2  \delta b_{\bar\jmath}+\omega_{\text{tr}}^2(\xi-1)\delta \theta_j =0 \\
&\dot{\delta \theta}_j+2 \dot{ \delta b_j}+\frac{1}{2\tau} (\delta \theta_j-\delta \theta_{\bar\jmath})=0
\end{align}
with ${\bar\jmath}=x$ if $j=y$, and $ {\bar \jmath} =y$ if $j=x$.
The collective modes have solutions of the form $ \delta b_i(t)=b_i^0 e^{i \omega t}$ and $ \delta \theta_i(t)=\theta_i^0 e^{i \omega t}$. Substituting into the above equations, we obtain a set of four linear equations for $b_i^0$ and $\theta_i^0$, which have non-zero solutions when the determinant of the associated matrix is zero. This gives a polynomial equation in $\omega$ that can be written 

\begin{equation}
\omega^2(\omega^2-\omega_{\text{Br}}^2)\left[ \left(\omega^2-\omega_{\text{cl}}^2\right)  -\frac{i}{\omega \tau}\left( \omega^2-\omega_{\text{hd}}^2\right) \right]=0,
\label{eq:coll-modes-sols}
\end{equation}
with
$ \omega_{\text{Br}} = 2 \omega_{\text{tr}}$,
$\omega_{\text{hd}} = \sqrt{2} \omega_{\text{tr}}, $ and $
\omega_{\text{cl}} =  \sqrt{2(2-\xi)} \omega_{\text{tr}}$. 

The solution $\omega=\omega_{\text{Br}}$ to Eq.~\eqref{eq:coll-modes-sols} corresponds to the breathing mode. It is purely real, and it is independent of $\xi$ and $\tau$, and hence independent of all system parameters other than $\omega_{\text{tr}}$.  As mentioned  this is a consequence of the scaling symmetry of the system. In the future it will be interesting to investigate the effects of the breakdown of this symmetry due to quantum effects.

\begin{figure}
 \includegraphics[scale=.5]{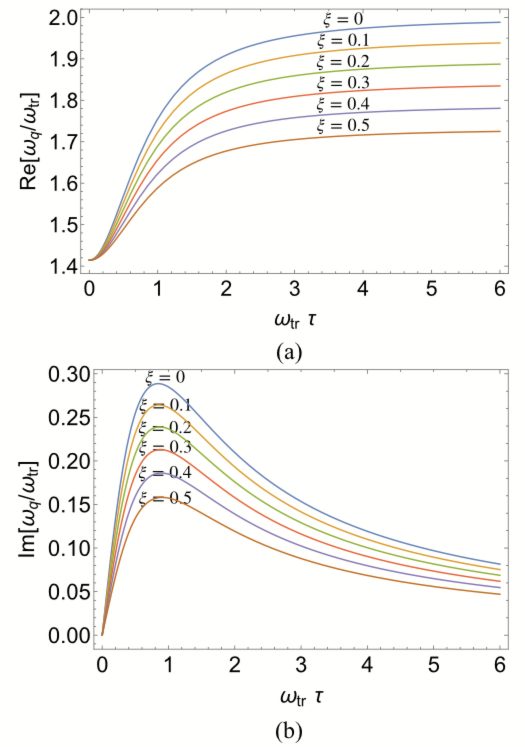}
 \vspace{-0.1in}
 \caption{Scaled quadrupole mode frequency($ \frac{\omega_q}{\omega_{\text{tr}}}$) vs scaled scattering time $\omega_{\text{tr}}\tau$. (a) Real part of the mode frequency vs scattering time, (b) imaginary part of the mode frequency vs scattering time. \label{fig:quad-modes}}
 \end{figure}

 The  solutions to the term in brackets in Eq.~\eqref{eq:coll-modes-sols}, 
$\left(\omega^2-2(2-\xi)\omega_{\text{tr}}^2\right)  -\frac{i}{\omega \tau}\left( \omega^2-2 \omega_{\text{tr}}^2\right) =0
$, give the quadrupole modes' (complex) resonance frequencies. In the hydrodynamic limit, $\omega_{\text{tr}}\tau \rightarrow 0$, the solution is $\omega=\omega_{\text{hd}}=\sqrt{2}\omega_{\text{tr}}$, and in the collisionless limit, $\omega_{\text{tr}} \tau \rightarrow \infty$, the solution is $\omega=\omega_{\text{cl}}=\sqrt{2(2-\xi)}\omega_{\text{tr}}$.  The real part of the frequency smoothly crosses over between these two limits [see Fig.~\ref{fig:quad-modes}(a)] while the imaginary part is zero in these two limits, peaking in between [see Fig.~\ref{fig:quad-modes}(b)]. 

This behavior of the imaginary part of $\omega$ is very general, and can be understood using the following argument. In the hydrodynamic limit  $\omega \tau \ll 1$, there is no dissipation since  frequent collisions force the deviations from local thermodynamic equilibrium, which are necessary for dissipation,  to be negligible \cite{pethick2008bose}. In this limit, the collective mode frequency is:
\begin{equation}
  \omega = \omega_{\text hd} - i \frac{\tau (\omega_{\text cl}^2 - \omega_{\text hd}^2)}{2}  .
\end{equation}  
On the other hand, in the collisionless limit $\omega \tau \gg 1$, there are few collisions per oscillation period, so dissipation is again negligible. In this limit, the collective mode frequency is:
\begin{equation}
  \omega = \omega_{\text cl} - i \frac{\omega_{\text cl}^2 - \omega_{\text hd}^2}{2 \tau \omega_{\text cl}^2}  .
\end{equation} 
It is clear that the damping rate must peak somewhere between the collisionless and hydrodynamic limits. Numerically, we find that the peak occurs, when $\omega \tau \sim 1$.

As detailed in Appendix~\ref{sec:xi-calc}, $\xi$ is given by
\begin{equation}
    \xi = \frac{g_{\text{2D}}(N-1)}{2 \frac{\pi \hbar^2}{m}} F(\frac{T}{\hbar\omega_{\text{tr}}},\frac{N_p}{N}),
\end{equation}
where $F(\frac{T}{\hbar\omega_{\text{tr}}},\frac{N_p}{N})$ is defined as the ratio of one-dimensional integrals in Eq.~\eqref{xi full form} and in general can  be evaluated numerically. In this calculation, we have ignored the mean-field contribution to the equilibrium distribution function, $f^0$. This approximation gives the value of $\xi$ to the leading order in $g_{\text{2D}}$. It is possible to evaluate $\xi$ analytically in the low temperature and high temperature limits. In the low temperature limit, when  $T \ll T_F$,  $\xi$ is 
\begin{equation}
\xi=\frac{g_{2D} m}{2 \pi \hbar^2}(N-1).
\end{equation}
In the high temperature limit, when  $T \gg T_F$,  $\xi$ is 
\begin{equation}
\xi=\frac{1}{8\pi}\frac{g_{\text{2D}}(N-1)}{\frac{\hbar^2}{m}}\frac{N_p}{N} \left(\frac{\hbar\omega_{\text{tr}}}{T}\right)^2.
\end{equation}

It is interesting to note that commonly-employed alternative approaches to computing the collective mode frequencies for a 2D SU($2$) Fermi gas have neglected the mean-field contribution~\cite{baur:collective_2013,Chiacchiera:damping_2013}. Due to this approximation, neither the breathing mode, nor the quadrupole modes show any mean-field shift in these calculations. Moreover, including the mean-field contribution in the Boltzmann equation in these approaches  leads to an unphysical shift in the breathing mode frequency due to the improper treatment of the SO(2,1) symmetry. As shown in the recent experiment by He {\it et al.}~\cite{he:collective_2019}, accounting for the mean-field shifts in the quadrupole mode can be important, especially as  $N$ increases. Our approach captures the mean-field effects while not inducing an unphysical shift in the breathing mode frequency.

 \subsection{Damping Rate Calculation \label{sec:-compute-xi-tau}}
In this section, we outline our calculations for the relaxation time $\tau$, which employs a few common approximations.
Following Ref.~\cite{baur:collective_2013}, we write $f({\bf r, p}) = f_0({\bf r, p}) + \left[f_0({\bf r, p})(1-f_0({\bf r, p}))\right] \phi({\bf r,p})$. Rearranging and taking moments of Eq.~\eqref{relaxation time apprx}, the relaxation time is given by 
\begin{equation}
    \frac{1}{\tau}=\frac{\langle \phi^* I_{\text{coll}}[\phi] \rangle}{\langle |\phi|^2f^0(1-f^0)\rangle}.
\end{equation}

To evaluate the relaxation rate, we use the following ansatz for $\phi({\bf r, p})$:
\begin{equation}
    \phi = p_x^2-p_y^2. \label{eq:phi-ansatz}
\end{equation}
The linearized collision integral then reads:
\begin{eqnarray}
    I_{\text{coll}} [f_1]=&&\int\!\frac{d^2 p_2}{(2 \pi \hbar)^2}\frac{m \hbar}{4 \pi} \int_0^{2 \pi}  \! d\theta |\mathcal{T}|^2   (\phi_1+\phi_2-\phi_3-\phi_4)\nonumber \\
    &&\hspace{0.2in}{}\times f_1 f_2 (1-f_3)(1-f_4) (N-1)
\end{eqnarray}

The calculations are easiest in the high temperature limit, when the Pauli blocking factors can be ignored. For weakly interacting Fermi gases, this approach overestimates the damping rate (by about $50 \%$) for typical experimental temperatures  ($T\sim 0.5 T_F$) \cite{baur:collective_2013}. We note that even in an anisotropic trap, this approximation gives a lower bound on the relaxation time in the weakly interacting limit. 

A more accurate estimate of the damping rate can be obtained by accounting for the Pauli blocking. This approach involves solving a six dimensional integral numerically, and overestimates the damping rate by about $10\%$ for $N=2$. It is possible to further systematically improve the damping rate estimate, by using ansatz for the function $\phi$ that are more flexible than Eq.~\eqref{eq:phi-ansatz}. Using a sufficiently complete basis, the result will converge to the true damping rate (within the relaxation time approximation).  The damping rate obtained using this technique is always bounded from below by the actual damping rate \cite{pethick2008bose,enss:shear_2012}, so convergence may be monitored as the damping tends to its minimum, in a manner analogous to other variational calculations, e.g. the convergence of the energy to its minimum when  calculating the quantum mechanical ground state energy. 

The relaxation time in this high temperature limit is found to be
\begin{eqnarray}
\label{tau eqn}
    \frac{1}{\tau} 
    &=&  \frac{ N-1 }{\tau_0}
\end{eqnarray}
where
\begin{equation} 
\frac{1}{\tau_0}=\frac{\pi N_p (\hbar \omega_{\text{tr}})^2}{2 N \hbar  k_B T} G\left(\frac{\hbar^2}{m a_{\rm 2D}^2 k_B T }\right)    
\end{equation}

and
\begin{equation}
    G(x) = \int_0^{\infty} dz \frac{z^2 \exp(-z)}{\ln{(x/z)}^2 + \pi^2}.
\end{equation}

\subsection{Experimental implications \label{sec:experiment}}
In this subsection, we compute the values of the mean-field shifts of the quadrupole modes and the damping rates for realistic experimental parameters.  Following He~{\it et al.} \cite{he:collective_2019}, we use the parameters $T=60\,\text{nK}$, $T/T_F=0.42$, $\omega_z=2\pi \times 59\,$kHz, $\omega_{\text{tr}}=2 \pi\times 185\,$Hz, $N_p/N=100$, $T/\hbar \omega_{\text{tr}}=5.94$, $\ln(k_F a_{\text{2D}})=-4.3$ to estimate the quadrupole mode shifts and the damping rates. For these parameters, the interaction strength is
\begin{equation}
\frac{g_{\text{2D}}}{\frac{\hbar^2}{m}}= - \frac{2 \pi}{\ln(k_F  a_{\text{2D}})}=1.461.
\end{equation}
Using Eq.(\ref{xi full form}), $\xi$, the parameter controlling the mean field shift,   is then
\begin{equation}
    \xi= \frac{g_{\text{2D}}}{2 \pi \frac{\hbar^2}{m}} (N-1) F(5.94,100)\approx 0.093(N-1)
\end{equation}

We compute the damping rate using Eq.~(\ref{tau eqn}), finding
\begin{equation}
    \frac{1}{\tau} \approx 0.095 (N-1) \frac{k_B T}{\hbar}.
\end{equation}

\begin{figure}
    \centering
    \includegraphics[scale=0.5]{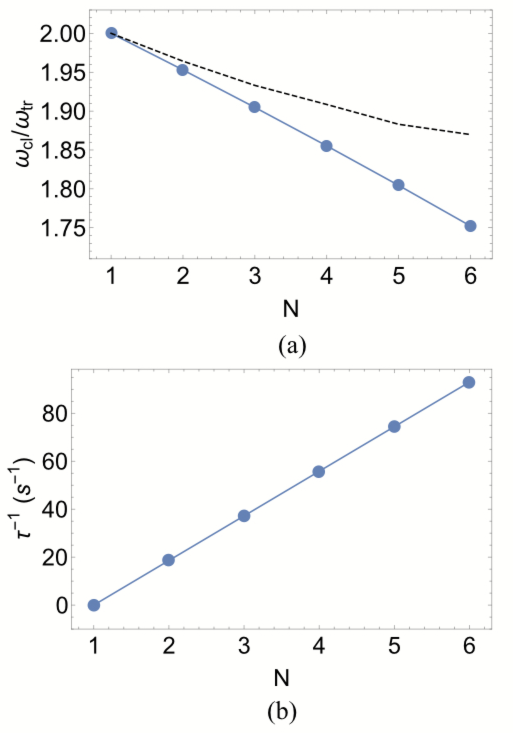}
    \caption{ (a) The ratio of the quadrupole mode frequency to the trap frequency in the collisionless limit. The solid blue line, and the dashed black line show the value of this ratio to exactly leading order in $g_{\text 2D}$ (neglecting the mean-field changes of the equilibrium distribution function $f^0$), and including the effects of mean-field interactions in $f^0$,  respectively. (b) The damping rate of the quadrupole modes, $\tau^{-1}$ (in units of $s^{-1}$) for parameters corresponding to the experiment in Ref. \cite{he:collective_2019}}
    \label{fig:shift-and-damping}
\end{figure}

 Fig.~\ref{fig:shift-and-damping} shows the collective mode frequency shifts and damping for the experimental parameters of Ref.~\cite{he:collective_2019} as a function of $N=1\ldots 6$. Our results for both the shifts and damping agree qualitatively with the experimental data presented in Ref. \cite{he:collective_2019}. Our prediction of the frequency shift is close to the experimentally observed frequency shift when $N=2$. We overestimate the frequency shift when $N\ge 3$, and our prediction for $\omega_{\text cl}$ is about $10\%$ less than the experimentally observed value when $N=6$. This discrepancy is likely due to our approximation of retaining fully only the leading order contribution of $g_{\text 2D}$ to the frequency shift, by neglecting the mean-field effects in the equilibrium distribution function. For larger $N$, a better estimate of $\xi$ can be obtained by including these mean-field effects  \cite{he:collective_2019}. Our damping rate estimate shows an  increase in damping rate with $N$, as observed experimentally. It is about $50 \%$ larger than the observed damping rate when $N=2$. This is in agreement with previous calculations \cite{baur:collective_2013}, and is in reasonable agreement with experiment after accounting for uncertainties in experimental parameters. However, for $N=6$, our estimate of the damping rate is roughly a factor of four larger than the measurements. It will be interesting to explore in future work to what extent this discrepancy results from  uncertainties in experimental parameters and to what extent it is from approximations employed in the theory. 

The presence of an anisotropic trap generally leads to a coupling of the breathing and quadrupole modes \cite{baur:collective_2013}. This leads to a decay of the breathing mode. However, the anisotropy in recent experiments is negligible (less than $1\%$) \cite{he:collective_2019}, and therefore we do not consider the effects of anisotropy in this paper.
 
 \subsection{3D results \label{sec:3D}} 
 
 The previous subsections computed the frequencies and damping rates of the collective modes of the 2D SU($N$) Fermi gas. We now briefly discuss the situation when there is no potential to confine the system to be two dimensional (e.g. no lattice), and only a weak trapping potential  in that direction. For our calculations, we assume that the gas is confined in a cylindrically symmetric harmonic potential, where $\omega_x=\omega_y=\omega_{\perp}$, and $\omega_z = \lambda \omega_{\perp}$. As outlined in Ref.~\cite{pedri2003dynamics}, the quadrupole mode frequencies are determined by the equation:
\begin{equation}
(\omega^2 - \omega_{\rm cl}^2) + \frac{i}{\omega \tau} (\omega^2-\omega_{\rm hd}^2) = 0,
\end{equation}
where $\omega_{\rm cl}^2 = 2 \omega_{\perp}^2 (2-\xi)$ and $\omega_{\rm hd}^2 = 2 \omega_{\perp}^2$, where the frequency shift $\xi$ is given by:
\begin{equation}
    \xi = g(N-1) \frac{\langle n(r) \rangle}{\langle 2 m \omega_i^2 r_i^2\rangle} = \frac{g(N-1)}{N} \frac{3 \langle n^{\rm tot}(r) \rangle}{\langle 2 m \omega_{\perp}^2 \rho^2\rangle},
\end{equation}
where $\rho = \sqrt{x^2 + y^2 + \lambda^2 z^2}$. We compute the collective mode frequencies in the low-temperature limit, where analytical results can be obtained. At zero temperature, the total density $n^{\rm tot}(r)$ can be approximated to be:
\begin{equation}
n^{\rm tot}(r) = \frac{8 N_p \lambda}{ \pi^2 R_F^3} (1- \frac{\rho^2}{R_F^2})^{3/2} \Theta(R_F-r)
\end{equation}
where $R_F= (48 N_p \lambda)^{1/6} \sqrt{\hbar/(m \omega_{\perp})}$ \cite{butts1997trapped}. Thus the shift, $\xi$ is given by:
\begin{eqnarray}
\xi &=& \frac{g(N-1)}{N} \frac{12 N_p \lambda}{ m \omega_{\perp}^2 \pi^2 R_F^3} \frac{\int d^3r (1- \frac{\rho^2}{R_F^2})^3 \Theta(R_F-r)}{\int d^3r \rho^2 (1- \frac{\rho^2}{R_F^2})^{3/2} \Theta(R_F-r)}  \nonumber \\
&=& \frac{g (N-1)}{m \omega_{\perp}^2 R_F^2 N}\frac{\sqrt{3 N_p \lambda (m \omega_{\perp})^3}}{\pi^2\hbar^{3/2}} \frac{4096}{945 \pi}.
\end{eqnarray}

 \section{Conclusions \label{sec:conclusions}}
 
We have calculated the collective modes for density oscillations of a harmonically-trapped 2D SU($N$) Fermi gas, carefully incorporating the SO(2,1) scaling symmetry. We employed a Boltzmann-Vlasov equation, which is valid for weak interactions and when the system is large compared to spatial coherences. We treated the collisions within the relaxation time approximation, which, while an uncontrolled approximation, is standard and captures the essential features of the hydrodynamic (and collisionless) limits. We solved this using a scaling ansatz, Eq.~\eqref{scaling}, for the semiclassical distribution function $f(\mathbf{r},\mathbf{p})$. In contrast to other methods, for example those based on the method of moments~
\cite{guery-odelin:collective_1999,ghosh:collective_2000, bruun:shear_2007,chiacchiera:role_2011},  the scaling ansatz preserves the SO(2,1) scaling symmetry, while also being flexible enough to allow for mean-field shifts. 

We have shown that the magnitude of the shifts are sufficiently large to measure if the system is confined in one dimension by an optical lattice. This suggestion is borne out in  recent experiments that were carried out as this manuscript was being finalized~\cite{he:collective_2019}. We note that in the collisionless limit, our calculations of shifts are equivalent to theirs, but ours apply across the full collisionless-hydrodynamic crossover, and also capture the mode damping.

The predicted quadrupole mode shifts agree with Ref.~\cite{he:collective_2019}'s experimental measurements after accounting for uncertainty in the experimental parameters. The predicted damping rate agrees with the order of magnitude observed in the experiments. For $N=2$, the damping rates agree with about $30\%$ relative error. However, for $N=6$ the predicted damping rate is roughly four times the experimentally measured rate.  It is possible that this arises from uncertainties in experimental parameters, which lead to a significant uncertainty already in the  collective mode shift frequencies. It's also possible that it arises from the approximations inherent in deriving and solving the BV equation. An interesting future direction will be precision experiments and calculations to pinpoint the reason for the discrepancy. 
 
Besides calculating the weakly interacting gases' collective modes, the results in this paper lay the groundwork for several future directions with ultracold SU($N$) gases collective modes and transport. One  direction is to explore strongly interacting gases, for instance to understand the shear viscosity~\cite{enss:shear_2012} or search for novel Fermi liquid behaviors and instabilities~\cite{yip:theory_2014,cheng:SUN_2017,how:nonanalytic_2018}. A strongly interacting regime may be achieved by tighter transverse confinement, higher densities, or optical Feshbach resonance~\cite{blatt:measurement_2011,yan:controlling_2013,nicholson:optical_2015}. In strongly interacting gases, the scaling anomaly should cause measurable effects, especially collective mode frequency shifts, and the ability to tune $N$ will shed new light into its effects. In this regard, it will be interesting to employ  other, sensitive probes of the scale anomaly, for example the momentum-space dynamics~\cite{murthy:quantum_2018}. 
It will also be interesting to study physics at lower temperatures, especially when the system becomes superfluid.
   Although the present calculations are performed for weakly interacting normal gases, such calculations form an important point of comparison for  strongly interacting and superfluid gases. 

Another direction would be to explore the spin modes~\cite{enss:quantum_2012,enss:transverse_2013}, which could allow a new window into strongly correlated spin transport.  These modes have been explored in $N=2$ alkali gases~\cite{koschorreck:universal_2013,bardon:transverse_2014,trotzky:observation_2015,valtolina:exploring_2017,enss:universal_2019}, but the potential spin structures are  richer for larger $N$. A scaling ansatz similar in spirit to our approach, with a spin dependent distribution function $f_{\sigma}({\bf r}, {\bf p}, t)$ has already been used to study spin dipole modes for $SU(2)$ fermi gases \cite{vichi1999collective,sommer2011universal}. Furthermore, it may be possible to generalize this scaling ansatz to include coherences. Thus, we expect that this technique can shed light on the spin collective modes of $SU(N)$ gases as well.

 \acknowledgements 
 We thank Stefan Natu and Erich Mueller for conversations. 
This material is based upon work supported with funds
from the Welch Foundation  Grant no.  C-1872 and from NSF Grant Nos.~PHY-1848304 and PHY-1607665.  K.~R.~A.~H thanks the Aspen Center for Physics, which is supported
by the National Science Foundation grant PHY-1066293,
for its hospitality while part of this work was performed.

\appendix

\section{Calculation of $\xi$ \label{sec:xi-calc}}

Using Eqs.~(\ref{MF potential}) and (\ref{shifteqn}) we have
\begin{equation}
\label{xi form}
\xi=\frac{g_{\text{2D}}(N-1)}{ m \omega_{\text{tr}}^2 N}\frac{\int d^2\mathbf{r}n^{\text{tot}}(\mathbf{r})^2}{\int d^2\mathbf{r}n^{\text{tot}}(\mathbf{r})r^2}.
\end{equation}
To calculate $\xi$, we need to know the form of equilibrium spatial density $n^{\text{tot}}(\mathbf{r})$ at a chemical potential set to match the total number of particles given by Eq.~(\ref{total particle}):
\begin{multline}
\label{np equation}
N_p=\int d^2 \mathbf{r} n^{\text{tot}}(\mathbf{r})=N\int d^2 \mathbf{r} \int \frac{d^2 \mathbf{p}}{\left(2 \pi\right)^2} f^0(\mathbf{r},\mathbf{p},\mu,T),
\end{multline}
where 
\begin{equation}
f^0=\frac{1}{e^{\left(\frac{p^2}{2 m}+V(\mathbf{r}+U_{\text MF})-\mu \right )/T}+1}.
\end{equation}
The chemical potential $\mu$ is determined by fixing $N_p = \int d^2 {\bf r} n^{\text{tot}}(\mathbf{r})$. We can then numerically evaluate $\xi$ in Eq.~(\ref{xi form}) for the total particle number $N_p$ at a temperature $T$, and trap frequency $\omega$. The mean-field contribution to $f^0$ can be ignored in the weakly interacting regime, since it only changes the density perturbatively. In the following analysis, we set $U_{\text MF} = 0$.
One can show that Eq.~(\ref{np equation}) can be written as,
 \begin{multline}
 \label{mu scaling}
 \frac{N_p}{N}\left(\hbar\omega_{\text{tr}} \right)^2 =\int \frac{dv du}{e^{\beta(v+u-\mu)}+1}=-\frac{\operatorname{Li}_2(-e^{\beta \mu})}{\beta^2}
\end{multline}
where $\operatorname{Li}_\alpha(z)$ is the polylog function.\\

In general this equation can be solved numerically, as discussed in Sec.~\ref{sec:scaling-app}. In the next subsections we analytically solve it in the high- and low-temperature limits.

\subsection{Low Temperature limit $T \ll \hbar\omega_{\text{tr}}\sqrt{2 \frac{N_p}{N}}$}

For $T \rightarrow 0$, we have
\begin{align}
\frac{N_p}{N}\left(\hbar\omega_{\text{tr}} \right)^2 &=\int_0^{\infty} \!dv \int _0^{\infty} \! du  \, \Theta(\mu-(v+u)) \\
&=\frac{\mu(T=0)^2}{2}\\
&=\frac{E_F^2}{2},
\end{align}
where $\Theta$ is the Heaviside theta function and $E_F$ is the Fermi energy, \begin{equation}
E_F= \hbar\omega_{\text{tr}} \sqrt{2\frac{N_p}{N}}.
\end{equation}
The density is
\begin{multline}
n^{\text{tot}}(\mathbf{r})=N  \int \frac{d^2 \mathbf{k}}{\left(2 \pi\right)^2} f^0=\frac{N}{2\pi}\frac{m}{\hbar^2}\int ds \frac{1}{e^{\beta(s+V(\mathbf{r})-\mu}+1},
\end{multline}
which at low temperature is
\begin{equation}
n^{\text{tot}}(\mathbf{r})=\frac{N}{2\pi}\frac{m}{\hbar^2} \left(E_F- V(\mathbf{r})\right) \Theta \left(E_F- V(\mathbf{r}) \right).
\end{equation}
Using the expression for $E_F$ and $V(\mathbf{r})$, we get, 
\begin{equation}
\label{low T density}
n^{\text{tot}}(\mathbf{r})=\frac{N}{4 \pi}\frac{r_0^2-r^2}{l^4}\Theta(r_0-r)
\end{equation}
where $r_0=l\left( 8\frac{N_p}{N}\right)^{\frac{1}{4}}$ and $l=\sqrt{\frac{\hbar}{m \omega_{\text{tr}}}}$. 
Using Eqs. (\ref{low T density}) and (\ref{xi form}), at low temperature($T \ll E_F$) we find
\begin{equation}
\xi= \frac{g_{\text{2D}}(N-1)}{\frac{2 \pi \hbar^2}{m}}.
\end{equation}

\subsection{High temperature limit $T \gg  \hbar\omega_{\text{tr}}\frac{N_p}{N} $}
At high temperature, \begin{align}
\frac{N_p}{N}\left(\hbar\omega_{\text{tr}} \right)^2& =\int \!\frac{dv du}{e^{\beta(v+u-\mu)}+1} \\
&\approx \int \! dv du \, e^{\beta(\mu-v-u)}\\
&=\frac{1}{\beta^2}e^{\beta \mu} 
\end{align}
so 
\begin{equation}
     \mu=T \log \left(\frac{N_p}{N} \left( \frac{\hbar\omega_{\text{tr}}}{T}\right)^2 \right).
\end{equation}
This approximation is only true when $e^{-\beta \mu} \gg 1 \rightarrow T \gg  \hbar\omega_{\text{tr}}\frac{N_p}{N} $.
With this approximation, at high temperature density becomes, \begin{equation}
n^{\text{tot}}(\mathbf{r})=\frac{N_p}{2 \pi}\frac{\hbar\omega_{\text{tr}}}{T}\frac{e^{-\beta V(\mathbf{r})}}{l^2}
\end{equation}
and, 
\begin{equation}
\xi= \frac{1}{8\pi}\frac{g_{\text{2D}}(N-1)}{\frac{\hbar^2}{m}}\frac{N_p}{N} \left(\frac{\hbar\omega_{\text{tr}}}{T}\right)^2.
\end{equation}

\subsection{Scaling analysis\label{sec:scaling-app}}

Eq.~(\ref{mu scaling}) implies that  $\tilde{\mu}=S(\frac{N_p}{N},\tilde{T})$, where $\tilde{\mu}\equiv \frac{\mu}{\hbar\omega_{\text{tr}}}$, $\tilde{T} \equiv \frac{T}{\hbar\omega_{\text{tr}}}$, and  
$S=-\tilde{T}\log[\operatorname{Li}_2^{-1}[-\frac{N_p}{N} \frac{1}{\tilde{T}^2}]]$.
The density is 
\begin{multline}
n^{\text{tot}}(\mathbf{r})=\frac{N}{2\pi}\frac{m}{\hbar^2}\int ds \frac{1}{e^{\beta(s+V(\mathbf{r})-\mu}+1}\\
=\frac{N}{2\pi}\frac{1}{l^2} \int d\tilde{s}\frac{1}{e^{\frac{\tilde{s}}{\tilde{T}}}e^{\left(\frac{r}{l}\right)^2\frac{1}{2 \tilde{T}}}e^{-\frac{\tilde{\mu}}{\tilde{T}}}+1} = \frac{N}{2 \pi l^2} h(\frac{r}{l},\tilde{T},\frac{N_p}{N}),
\end{multline}
implicitly defining $h$ on the last line. Scaling Eq.~(\ref{xi form}), $\xi$ is given by 
\begin{align}
\label{xi full form}
\xi&=\frac{g_{\text{2D}}(N-1)}{2 \pi m \omega_{\text{tr}}^2l^4}\frac{\int d^2\mathbf{r}h(r,\tilde{T},\frac{N_p}{N})^2}{\int d^2\mathbf{r}r^2h(r,\tilde{T},\frac{N_p}{N})}\\ 
&= \frac{g_{\text{2D}}(N-1)}{\frac{2 \pi \hbar^2}{m}} F(\tilde{T},\frac{N_p}{N}) \\
&=\xi_0 F(\tilde{T},\frac{N_p}{N}).
\end{align}
Here $F$ is implicitly defined in the second line, and $\xi_0=\frac{g_{\text{2D}}(N-1)}{\frac{\pi \hbar^2}{m}}$. One can   numerically calculate $F(\tilde{T},\frac{N_p}{N})$ in general, while the prior two sections show   that $F(0,\frac{N_p}{N})=1$ and $F(\tilde{T} \gg \frac{N_p}{N},\frac{N_p}{N})=\frac{1}{4}\frac{N_p}{N}\frac{1}{\tilde{T}^2}$.

\begin{figure}[h]
\includegraphics[scale=0.45]{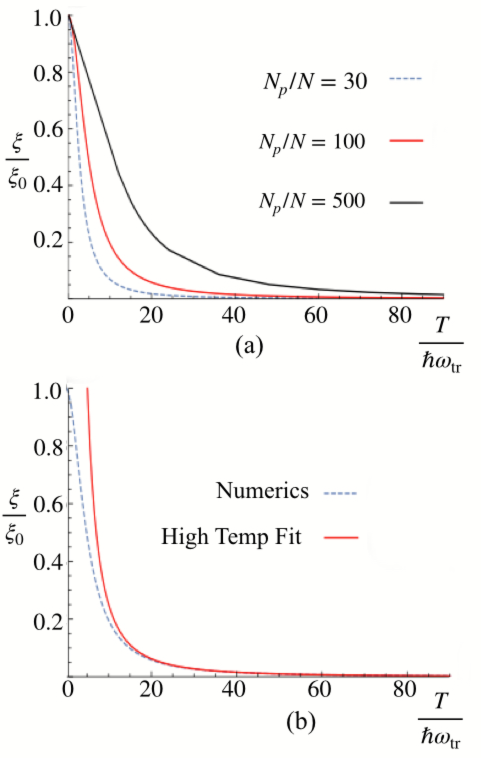}
\caption{(a)$\frac{\xi}{\xi_0}$ vs $\frac{T}{\hbar\omega_{\text{tr}}}$ for different values of $\frac{N_p}{N}$ and (b) Numerical calculation for $\frac{\xi}{\xi_0}$ vs theoretical calculation at high temperature limit for $\frac{N_p}{N}=100$.}
\end{figure}

\end{document}